\documentstyle[twocolumn]{jpsj}

\def\ggs{\buildrel\textstyle > \over {\hbox{\raise0.2ex\hbox{$\sim$}}}}
\def\lls{\buildrel\textstyle < \over {\hbox{\raise0.2ex\hbox{$\sim$}}}}
\def\gsim{\,\lower0.75ex\hbox{$\ggs$}\,}
\def\lsim{\,\lower0.75ex\hbox{$\lls$}\,}

\newcommand{\Js}{J_{\rm s}}

\newcommand{\kb}{{\bf k}}
\newcommand{\qb}{{\bf q}}

\newcommand{\simg}{\stackrel{>}{_\sim}}
\newcommand{\siml}{\stackrel{<}{_\sim}}

\def\runtitle{Pseudogap Phenomena and Phase Diagram \\ in the 2-Band Hubbard Model}
\def\runauthor{Akito {\sc Kobayashi} {\it et al.}}

\title
{Pseudogap Phenomena and Phase Diagram \\ in the 2-Band Hubbard Model}

\author
{
Akito {\sc Kobayashi}$^{1}$, Atsushi {\sc Tsuruta}$^{1}$, 
Tamifusa {\sc Matsuura}$^1$ and Yoshihiro {\sc Kuroda}$^{1, 2}$, 
}

\inst
{
$^{1}$Department of Physics, Nagoya University, Nagoya 464-8602 \\
$^{2}$CREST, Japan Science and Technology Corporation (JST) \\
}

\recdate{
\hspace*{20mm}
}

\abst{
High-$T_{\rm c}$ superconducting materials (HTSC) have anomalous properties such as the pseudo-gap or spin-gap {\it etc.}, in Hall coefficient, $1/T_1T$ and the density of states {\it etc.}
First including effects of strong on-site repulsion between d-electrons at Cu-sites, we obtain quasi-particles with super-exchange interaction $J_{\rm s}$, whose band width tends to zero, {\it i.e.}, the system goes to insulator, as $\delta$ tends to zero.
The quasi-particles correspond to Zhang-Rice singlet states.
$J_{\rm s}$ larger than the band width combined with the 2-dimensional character of the system induces strong antiferro-magnetic (AF) and superconducting (SC) fluctuations.
We treat effects of the AF ones in the FLEX approximation and those of the SC ones in the self-consistent t-matrix approximation to show that both fluctuations in the under-doped region start to increase at $T_0$ as $T$ decreases from $T\gg T_{\rm c}$, the AF ones dominate the SC ones at $T>T_{\rm sg}$, while SC ones dominate at $T<T_{\rm sg}$.
This cross-over of the fluctuations causes the anomalous phenomena in the under-doped region.
We also obtain the $T-\delta$ phase diagram of HTSC consistent to one observed in experiments.
}

\kword{pseudogap, superconducting gap, superconductivity, superconducting fluctuation, spin fluctuation, superexchange interaction, Hubbard model} 

\begin{document}
\input epsf

\sloppy
\maketitle

Since the discovery of the high-$T_{\rm c}$ superconducting materials (HTSC), the anomalous metallic phase have attracted much attention, where various anomalous properties have been observed in considerably wide temperature range higher than the superconducting transition temperature $T_{\rm c}$\cite{Yasu,Regnault,Nishikawa,Sato,Renner,Loeser,Ding,Norman}.
The antiferromagnetic fluctuations (AF fluctuations) are observed in various experiments, {\it e.g.} the NMR\cite{Yasu}, the neutron scattering\cite{Regnault} and the Hall coefficients\cite{Nishikawa,Sato}.
On the other hand, the pseudogap is also observed in various experiments, {\it e.g.} the tunneling spectroscopy\cite{Renner} and the angle-resolved photo emission spectroscopy (ARPES)\cite{Loeser,Ding,Norman}, which is of the same size and has the same ${\bf k}$-dependence as the superconducting gap (SC gap).
There are several different views regarding the origin of the anomalous metallic phase, especially concerning the pseudogap.
For example, the pseudogap may be interpreted as a spin-excitation gap in the singlet RVB states obtained in the $t$-$J$ model\cite{Suzumura,Fukuyama}, or as precursor of the SC gap\cite{Emery,Levin,pgdp,Onoda,Yanase,Koikegami}.

Based on the above experimental results, we have suggested a mechanism of the superconductivity and the anomalous metallic phase\cite{pgdp}, where the AF fluctuations and the superconducting fluctuations (SC fluctuations) play essential roles as described in the followings.
These fluctuations are induced by significant features of the HTSC, {\it i.e.} a quasi-$2$ dimentional metal in the vicinity of the Mott insulator.
In metals near Mott insulators, doped carriers create narrow quasi-particle band inside the charge transfer gap called as in-gap states, which have band width proportional to the hole doping rate $\delta$ and interact with each other through the superexchange interaction $J_{\rm s}$\cite{Jichu,Ono2,Fukagawa}.
It is essentially the strong correlation effects.
Strong AF fluctuations are induced by $J_{\rm s}$ in the in-gap state.
The pairing interaction mediated by the AF fluctuations induces the high-$T_{\rm c}$ superconductivity with $d_{x^2-y^2}$ symmetry.
At the same time, the pairing interaction promotes strong SC fluctuations because of the quasi-2D electronic system, which induces the pseudogap in the quasi-particle spectrum.
The pseudogap suppresses the AF fluctuations.
Thus, effects of both of the AF fluctuations and the SC fluctuations need to be treated in a consistent way.

In our previous studies\cite{pgdp}, we have shown that the SC fluctuations induce the pseudogap and plays key roles in the anomalous metallic phase by using the self-consistent $t$-matrix approximation based on the in-gap state and $J_{\rm s}$, which are derived in the 2-band Hubbard model with the help of the slave boson technique together with the $1/N$-expansion theory\cite{Jichu,Ono2,Fukagawa}.
We have shown that $1/T_1T$ of the NMR increases as T decreases at high temperatures, and reaches a maximum at $T_{\rm sg}$ followed by a sharp drop in the under-doped region, due to the evolution of the pseudogap.
The evolution is consistent with those of the ARPES experiments.\cite{Norman}
It has been also shown that $T_{\rm c}$ is suppressed by the SC fluctuations in the under-doped region.

In the present study, we demonstrate the above senario explicitly by taking the AF fluctuations in the fluctuation-exchange (FLEX) approximation and the SC fluctuations in the self-consistent $t$-matrix approximation to treat both fluctuations on the equal footing, which are also based on the in-gap state and $J_{\rm s}$ in the 2-band Hubbard model.
By using the Green's functions thus obtained, we calculate $T_{\rm c}$, the antiferromagnetic transition temperature $T_{\rm N}$, $1/T_1T$, the spin susceptibility and the density of states.
As a result, we obtain the $T$-$\delta$ phase diagram which is consistent with that observed in HTSC.

First, we describe our model and formulation.
We take the simplest version of the $U=\infty$ 2-band Hubbard model for describing the electronic system in the ${\rm CuO}_2$ plane:
\begin{eqnarray} 
H &=& \varepsilon_p \sum_{\kb \sigma}c_{\kb \sigma}^+c_{\kb \sigma} 
       +  \varepsilon_d \sum_{i \sigma}d_{i\sigma}^+d_{i\sigma}
\nonumber \\       
&+& N_{\rm L}^{-\frac12}\sum_{i \kb \sigma}\{t_{i\kb}
           c_{\kb \sigma}^+d_{i\sigma}b_i^+ + h.c.\}, \label{hamiltonian}
\end{eqnarray}
which is treated within the physical subspace where local constraints
$\hat{Q}_i = \sum_\sigma d_{i\sigma}^+d_{i\sigma} + b_i^+b_i = 1$
hold\cite{Coleman,Jin}.
In the above, $c_{\kb \sigma}$, $d_{i\sigma}$ and $b_i$ are annihilation operators for a $p$-hole with wave vector ${\bf k}$ and spin $\sigma$, a $d$-hole at i-th site with spin $\sigma$ and a slave boson at i-th site, respectively, and $t_{i\kb} = t_\kb \exp(-i\kb\cdot{\bf R}_i)$, with
$t_{\kb} = 2t[1-\frac12 (\cos k_{x}a + \cos k_{y}a)]^\frac12$
,where $a$ is the lattice constant (we set $a=1$), and $N_{\rm L}$ is the total number of lattice sites.

The quasi-particle Green's functions of the leading order in the $1/N$-expansion are given by~ \cite{Jichu,Ono2}
\begin{eqnarray} \label{GK} 
G_0 (\kb, \omega ) = \sum_{\gamma=\pm}A_\gamma (\kb )/(\omega  - E_\gamma (\kb )+{\rm i}0^+ ),
\end{eqnarray}
with
$E_\gamma (\kb )=\frac12 [\varepsilon_{p}+\omega_0+\gamma ((\varepsilon_{p}-\omega_0)^2+4b t_{\bf k}^2 )^{1/2} ]$
,
$A_\gamma (\kb )=\gamma (E_\gamma (\kb ) - \omega_0)/(E_+ (\kb )- E_- (\kb ))$,
where $N$ represents the spin-orbital degeneracy of $d$-hole and then we set $N=2$ in the present case.
The binding energy $\omega_0$ and the residue $b$ of the pole in the slave-boson Green's function are self-consistently determined together with the chemical potential $\mu$.
Here, it is noted that the solution $E_{-} (\kb )$ denotes the in-gap state emerging inside the charge transfer gap, $\Delta \equiv \varepsilon_p - \varepsilon_d$, upon doping carriers to a Mott insulator.
The band width and the residue of the in-gap state roughly equal $2\omega_0$ and $b$, respectively, and are proportional to $\delta$ in the under-doped region.
The effects of the interactions among the in-gap states are of higher-order terms in the $1/N$-expansion.\cite{Miura,Hirashima,Azami3}
Recently, it has been shown that quasi-particle interactions via the superexchange interaction $J_{\rm s}$ play dominant roles in under-doped systems.~\cite{Fukagawa} 

Now, we derive the coupled equations of the AF fluctuations and the SC fluctuations.
The AF fluctuations via $J_{\rm s}$ are given by
\begin{equation}
V({\bf q}, \omega ) = J_{\rm s}({\bf q})/(1-J_{\rm s}({\bf q}) \widetilde{\chi}^{\rm s} ({\bf q}, \omega )),
\end{equation}
with
$J_{\rm s}({\bf q})=-J_{\rm s} (\cos (q_x )+\cos (q_y ))$
and the irreducible spin susceptibility
\begin{eqnarray}
\widetilde{\chi}^{\rm s} ({\bf q},\omega ) &=& N_{\rm L}^{-1}\sum_{\bf k} \Omega_{{\bf k}+{\bf q},{\bf k}} \pi^{-1}\int {\rm d}x f(x) \nonumber \\
&\times& [{\rm Im}G({\bf k}+{\bf q},x)G({\bf k},-\omega +x)^* \nonumber \\
&+& G({\bf k}+{\bf q},\omega +x){\rm Im}G ({\bf k},x) \nonumber \\
&-& {\rm Im}F({\bf k}+{\bf q},x)F({\bf k},-\omega +x)^* \nonumber \\
&-& F({\bf k}+{\bf q},\omega +x){\rm Im}F ({\bf k},x)],
\end{eqnarray}
where
$\Omega_{{\bf k},{\bf k}^\prime}=b^2 t_{\bf k}^2 t_{{\bf k}^\prime}^2/((E_- ({\bf k}) -\omega_0 )^2 (E_- ({\bf k}^\prime ) -\omega_0 )^2)$.

It was shown that a component with the $d_{x^2 -y^2}$ symmetry among various components of the spin-fluctuation-mediated interaction contributes dominantly to the pairing interaction.\cite{Azami3}
We take only the component with the $d_{x^2 -y^2}$ symmetry of the pairing interaction as
\begin{equation}
v_d =N_{\rm L}^{-2} \sum_{\bf k} \sum_{\bf k^\prime} \psi_{\bf k} V({\bf k}-{\bf k^\prime}, 0) \psi_{\bf k^\prime} ,
\end{equation}
with
$\psi_{\bf k} =\cos (k_x a)-\cos (k_y a)$.

The pairing susceptibility $\widetilde{\chi}^{\rm p} ({\bf q}, \omega )$, which is the irreducible part of the $t$-matrix, is given by\cite{Kosuge}
\begin{eqnarray}
\widetilde{\chi}^{\rm p} ({\bf q}, \omega ) &=& \widetilde{\chi}_{\rm n}^{\rm p} ({\bf q}, \omega ) \nonumber \\
&+& v_d \widetilde{\chi}_{\rm a}^{\rm p} ({\bf q},\omega ) \widetilde{\chi}_{\rm a}^{{\rm p} \dagger} ({\bf q},\omega )/(1-v_d \widetilde{\chi}_{\rm n}^{\rm p} ({\bf q},\omega )),
\end{eqnarray}
where the pairing susceptibility due to the normal Green's fucntions is given by
\begin{eqnarray}
&\widetilde{\chi}&_{\rm n}^{\rm p} ({\bf q},\omega )=-N_{\rm L}^{-1}\sum_{\bf k}\psi_{\bf k}^2 \Lambda_{{\bf k},{\bf q}-{\bf k}} \pi^{-1} \int {\rm d}x  \nonumber \\
&\times& [ f(x){\rm Im} G ({\bf k},x) G ({\bf q}-{\bf k},\omega -x) \nonumber \\
&-&f(-x) G ({\bf k},\omega -x) {\rm Im} G ({\bf q}-{\bf k},x)],
\end{eqnarray}
while that due to the anomalous Green's fucntions by
\begin{eqnarray}
&\widetilde{\chi}&_{\rm a}^{{\rm p} (\dagger )} ({\bf q},\omega )=-N_{\rm L}^{-1}\sum_{\bf k}\psi_{\bf k}^2 \Lambda_{{\bf k},{\bf q}-{\bf k}} \pi^{-1} \int {\rm d}x  \nonumber \\
&\times& [ f(x){\rm Im} F^{(\dagger )} ({\bf k},x) F^{(\dagger )} ({\bf q}-{\bf k},\omega -x) \nonumber \\
&-&f(-x) F^{(\dagger )} ({\bf k},\omega -x) {\rm Im} F^{(\dagger )} ({\bf q}-{\bf k},x)]
\end{eqnarray}
and
$\Lambda_{{\bf k},{\bf k}^\prime}=b^2 t_{\bf k}^2 t_{{\bf k}^\prime}^2/((\pi T)^2 +\omega_0^2 )^2$.
In the above both $\Omega_{{\bf k},{\bf k}^\prime}$ and $\Lambda_{{\bf k},{\bf k}^\prime}$ are the vertices connecting $p$- and $d$-holes.\cite{Fukagawa}

Self-energy corrections due to the SC fluctuations in the $t$-matrix approximation are given by
\begin{eqnarray}
\Sigma_{\rm SCf} ({\bf k},\omega )=\psi_{\bf k}^2 N_{\rm L}^{-1}\sum_{\bf q} \Lambda_{{\bf k},{\bf q}-{\bf k}} \pi^{-1} \int {\rm d}x \nonumber \\
\times [ f(x) T({\bf q},x+\omega ) {\rm Im} G ({\bf q}-{\bf k},x) \nonumber \\
 - g(x) {\rm Im} T({\bf q},x) G ({\bf q}-{\bf k}, x-\omega )^\ast ],
\end{eqnarray}
with the $t$-matrix of the SC fluctuations
\begin{eqnarray}
T({\bf q},\omega )=v_d^2 \widetilde{\chi}^{\rm p} ({\bf q},\omega )/(1-v_d \widetilde{\chi}^{\rm p} ({\bf q},\omega )),
\end{eqnarray}
where $g(E) = (\exp\{E/k_{\rm B}T\} - 1 )^{-1}$.

Self-energy corrections due to the AF fluctuations in the fluctuation exchange (FLEX) approximation are given by
\begin{eqnarray}
\Sigma_{\rm AFf} ({\bf k},\omega )=N_{\rm L}^{-1}\sum_{\bf q} \Lambda_{{\bf k},{\bf k}-{\bf q}} \pi^{-1} \int {\rm d}x \nonumber \\
\times [ g(x) {\rm Im} V({\bf q},x )  G({\bf k}-{\bf q}, \omega -x) \nonumber \\
 - f(-x) V({\bf q},\omega -x) {\rm Im} G ({\bf k}-{\bf q}, x ) ].
\end{eqnarray}

The renormalized Green's function is given by
\begin{eqnarray}
G ({\bf k},\omega )=[G_{\rm 0} ({\bf k},\omega )^{-1} -\Sigma_{\rm SC} ({\bf k},\omega ) \nonumber \\
-\Sigma_{\rm SCf} ({\bf k},\omega ) -\Sigma_{\rm AFf} ({\bf k},\omega )]^{-1},
\end{eqnarray}
and the anomalous Green's function is
\begin{eqnarray}
F ({\bf k},\omega )=-G_{\rm n} ({\bf k},-\omega )^\ast \Delta_{\rm SC} ({\bf k}) G ({\bf k},\omega ),
\end{eqnarray}
where the self-energy due to the SC gap
\begin{eqnarray}
\Sigma_{\rm SC} ({\bf k},\omega )=- b^{-2} G_{\rm n} ({\bf k},-\omega )^\ast \vert \Delta_{\rm SC} ({\bf k}) \vert^2,
\end{eqnarray}
the renormalized Green's function without $\Sigma_{\rm SC} ({\bf k},\omega )$ by
\begin{eqnarray}
G_{\rm n} ({\bf k},\omega )=[G_{\rm 0} ({\bf k},\omega )^{-1} \nonumber \\
-\Sigma_{\rm SCf} ({\bf k},\omega )-\Sigma_{\rm AFf} ({\bf k},\omega )]^{-1}
\end{eqnarray}
and the SC gap with the $d_{x^2 -y^2}$ symmetry by
$\Delta_{\rm SC} ({\bf k})=v_d \psi_{\bf k} t_{\bf k}^2 \widetilde{\Delta}_{\rm SC}$.
The order parameter $\widetilde{\Delta}_{\rm SC}$ is determined by
\begin{eqnarray}
1-\alpha =-v_d N_{\rm L}^{-1}\sum_{\bf k}\psi_{\bf k}^2 \Lambda_{{\bf k},{\bf k}} \pi^{-1} \int {\rm d}x \nonumber \\
\times [ f(x){\rm Im} G_n ({\bf k},x) G (-{\bf k}, -x) \nonumber \\
 -f(-x) G_n ({\bf k}, -x) {\rm Im} G (-{\bf k},x)].
\end{eqnarray}
Both the SC and AF fluctuations are self-consistently treated by solving the coupled equations (3)-(16).

At $T=T_{\rm c}$, eq. (16) corresponds to the Thouless criterion,
$1-v_d \widetilde{\chi}^{\rm p} ({\bf 0},0) - \alpha =0$,
where $\alpha$ represents degree of the $3$D effect in the quasi-$2$D electronic system.
In the present model $1-v_d \widetilde{\chi}^{\rm p} ({\bf 0},0)$ does not reach zero at finite temperatures by the effect of $\Sigma_{\rm SCf} ({\bf k},\omega )$, because we take a 2D electronic ststem in the calculation of the above coupled equations.
We take $\alpha =0.5$ in the present study so as to have finite $T_{\rm c}$ of $O(100{\rm K})$.

We define $T_{\rm N}$ as the temperature where $1-J_{\rm s}({\bf Q}) \widetilde{\chi}^{\rm s} ({\bf Q},0)-\beta =0$, where ${\bf Q}=(\pi ,\pi )$ and $\beta$ represents degree of the $3$D effect.
In the present model $1-J_{\rm s}({\bf Q}) \widetilde{\chi}^{\rm s} ({\bf Q},0)$ also does not reach zero at finite temperatures by the effect of $\Sigma_{\rm AFf} ({\bf k},\omega )$ in the 2D electronic ststem.
We take $\beta =0.02$ in the present study.

By using the solusions of the above coupled equations, we calculate the NMR relaxation rate, 
\begin{equation}
(T_1T)^{-1} =N_{\rm L}^{-1} \sum_q F_{ab}({\bf q}) \lim_{\omega \rightarrow 0} {\rm Im} \chi^{\rm s} ({\bf q}, \omega )/\omega,
\end{equation}
where
$\chi^{\rm s} ({\bf q}, \omega ) = \widetilde{\chi}^{\rm s} (\qb, \omega)/(1-J_{\rm s} ({\bf q}) \widetilde{\chi}^{\rm s} (\qb, \omega))$
and the factor $F_{ab}({\bf q})$ is given in Ref. 26.

In actual numerical calculations, throughout the present study, we set $2t = 1.0$ (which is of $O(1 {\rm eV})$ in real systems), $\Delta = 2.5$ and $\Js = 0.1$.
For an example, we have $\omega_0 =0.0780$ and $b=0.0515$ at $\delta =0.1$
The total number of discrete points taken for the $\qb$-summation over the 2D first Brillouin zone is 32 $\times$ 32.
The $\omega$-integral over the region from $-2\omega_0$ to $2\omega_0$ is replaced by the $\omega$-summation of $80$ discrete points.

Figure 1 shows the $T$-$\delta$ phase diagram obtained by solving the above coupled equations.
Both the AF fluctuations and the SC fluctuations begin to increase at $T \cong T_0$ as $T$ decreases.
The AF fluctuations dominate in the region of $T_{\rm sg} \siml T \siml T_0$, while the SC fluctuations dominate in the region of $T_{\rm c} < T \siml T_{\rm sg}$, where $T_0$ and $T_{\rm sg}$ increase monotonically when $\delta$ decreases.
This crossover behavior induces the anomalous $T$-dependence of $1/T_1T$ as discussed in Fig. 2.
As a result of the domination by the SC fluctuations in the under-doped region, $T_{\rm c}$ has a maximum at $\delta \cong 0.1$ and decreases as $\delta$ decreases.
The key factor for obtaining these reasonable features is the narrow in-gap state with strong $J_{\rm s}$, {\it i.e.} $J_{\rm s}/W \simg 1$ in the under-doped region, where the band-width $W$ is approximately proportional to $\delta$.
This phase diagram accounts for essential features of the phase diagram observed in HTSC~\cite{Sato}.


\begin{figure}
\def\epsfsize#1#2{0.4#1}
\centerline{
\epsfbox{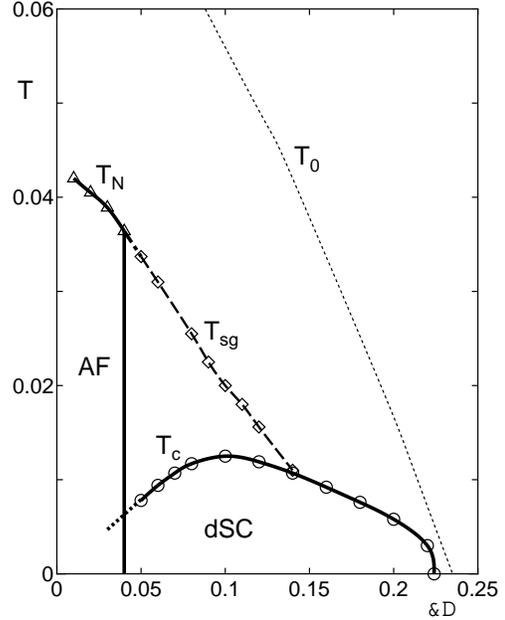}
}
\caption{
The $T$-$\delta$ phase diagram obtained by solving the coupled equations (3)-(16).
There is the superconducting state with $d_{x^2 -y^2}$ symmetry in the region of $T \le T_{\rm c}$.
The antiferromagnetic state exists in the region of $T \le T_{\rm N}$.
The spingap temperature $T_{\rm sg}$ is defined as the temperature at which $1/T_1T$ has a maximum as shown in Fig. 2.
We define $T_{\rm 0}$ as the temperature at which the AF fluctuations begin to be appreciable, where explicit values of $T_{\rm 0}$ are determined by the condition that $[\frac{\partial}{\partial T} (1/T_1 T) \vert_{T=T_{\rm 0}}]/[(1/T_1 T) \vert_{T=0.08, \delta =0.1}] =-0.02$.
We note that the SC fluctuations also become appreciable at $T \siml T_{\rm 0}$ as discussed in Fig. 3.
The metallic region at $T \siml T_{\rm 0}$ is the anomalous metallic phase.
}
\label{fig.1}
\end{figure}


Figure 2 shows $T$-dependences of $1/T_1T$ obtained by applying 3 different types of approximations for $\chi^{\rm s}$ in eq. (17) at $\delta=0.1$. 
In the case of RPA, $1/T_1T$ diverges at $T=T_{\rm N}^{\rm RPA}$, which indicates the system becomes the antiferromagnetic state.
In the case of FLEX, $1/T_1T$ does not diverge but increases monotonically as $T$ decreases.
By solving the coupled equations including both the AF fluctuations and the SC fluctuations, $1/T_1T$ is strongly suppressed in the low temperature region, and then has a maximum at $T_{\rm sg}$.
It is a typical example of the crossover behavior of the fluctuations in the anomalous phase.

\begin{figure}
\def\epsfsize#1#2{0.4#1}
\centerline{
\epsfbox{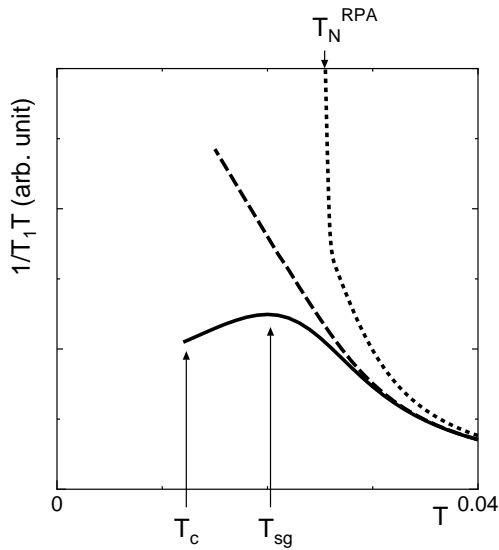}
}
\caption{
$T$-dependences of $1/T_1T$ obtained by applying 3 different types of approximations for $\chi^{\rm s}$ in eq. (17) at $\delta=0.1$.
The solid line represents $1/T_1T$ obtained by solving the coupled equations.
The dashed line represents $1/T_1T$ obtained by the FLEX approximation where only the self-energy corrections of the AF fluctuations are included.
The dotted line represents $1/T_1T$ obtained by the random phase approximation (RPA), {\it i.e.} no self-energy corrections of fluctuations are included.
}
\label{fig.2}
\end{figure}

Figure 3 shows $\omega$-dependences of the normalized density of states $\rho (\omega) /\rho^{\rm 0} (\omega_0)$ at $\delta=0.1$.
The pseudogap begins to open at $T=0.06 \cong T_0$ and develops gradually as $T$ decreases, suppressing the AF fluctuations, which induces the crossover of the fluctuations.
The pseudogap is induced by the SC fluctuations via the self-energy corrections, and has the same size and the same ${\bf k}$-dependence as the SC gap, {\it i.e.} the precursor of the SC gap.

\begin{figure}
\def\epsfsize#1#2{0.4#1}
\centerline{
\epsfbox{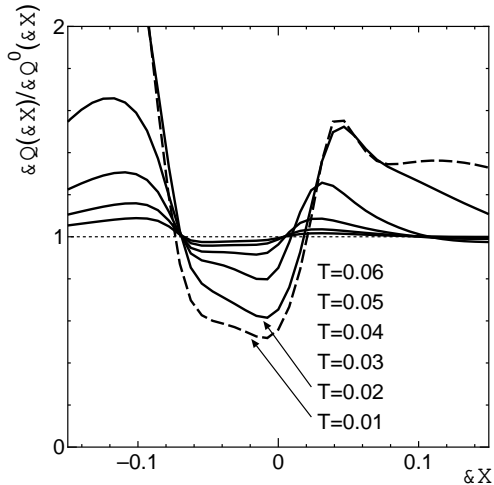}
}
\caption{
$\omega$-dependences of the normalized density of states $\rho (\omega) /\rho^{\rm 0} (\omega_0)$ at $\delta=0.1$ and $T=0.06$, $0.05$, $0.04$, $0.03$, $0.02$ and $0.01$, where $T_{\rm c}=0.012$.
The density of stats $\rho (\omega ) = -\frac{1}{\pi N_{\rm L}} \sum_{\bf k} {\rm Im} G({\bf k},\omega )$ is obtained by solving the coupled equations, and $\rho^{0} (\omega )$ is obtained by using the Green's function of the in-gap state (eq. (2)) where no self-energy corrections of fluctuations are included.
}
\label{fig.3}
\end{figure}

The present study sheds light on a new problem in strongly correlated systems where two kinds of strong fluctuations affect each other.
We have shown that the phase diagram of HTSC is described by our senario consistently.
The narrow in-gap state and strong $J_{\rm s}$ are the key factors for describing the features of the phase diagram in the under-doped region.
As for the quasi-particle interaction, the on-site Coulomb interaction is not effetive in the under-doped region, because it is reduced to same order of the band-width of the in-gap state, while $J_{\rm s}$ remains to be finite almost independent of $\delta$ in the under-doping region\cite{Fukagawa}.


The authors are grateful to Prof. M. Sato for fruitful discussions.
The present work has been partially supported by the Grant-in-Aid for Scientific Research from the Ministry of Education, Science, Sports and Culture, Japan.



\end{document}